\documentclass[twocolumn,aps,prl,superscriptaddress,footinbib,amsmath,amssymb,aps,floatfix]{revtex4-1}

\usepackage{bm}
\usepackage{graphicx}

\newcommand{\up}{|\!+\!z\rangle}
\newcommand{\down}{|\!-\!z\rangle}
\newcommand{\updown}{|\!\pm z\rangle}
\newcommand{\TTFi}{(T/T_F)_\mathrm{i}}
\newcommand{\svec}[1]{\bm{#1}}
\newcommand{\rvec}[1]{\vec{#1}}
\newcommand{\bk}{{\vec k}}
\newcommand{\bq}{{\vec q}}
\newcommand{\vzero}{{\vec 0}}
\newcommand{\M}{\bm{\mathcal{M}}}
\newcommand{\Jj}{\bm{\mathcal{J}}_{\!j}}
\newcommand{\grad}{\nabla_{\!j}}
\newcommand{\be}{\begin{equation}}
\newcommand{\ee}{\end{equation}}
\renewcommand{\Re}{\operatorname{Re}}

\newcommand{\T}{{\mathcal{T}}}

\begin{document}

\title{Observation of the Leggett-Rice effect in a unitary Fermi gas}
\author{S.\ Trotzky}
\author{S.\ Beattie}
\author{C.\ Luciuk}
\author{S.\ Smale}
\author{A.\ B.\ Bardon}
\affiliation{Department of Physics, University of Toronto, M5S 1A7 Canada}
\author{T.\ Enss}
\affiliation{Institut f\"ur Theoretische Physik, Universit\"at Heidelberg, D-69120 Germany}
\author{E.\ Taylor}
\affiliation{Department of Physics and Astronomy, McMaster University, L8S 4M1 Canada}
\author{S.\ Zhang}
\affiliation{Department of Physics, Center of Theoretical and Computational Physics, University of Hong Kong, China}
\author{J.\ H.\ Thywissen}
\affiliation{Department of Physics, University of Toronto, M5S 1A7 Canada}
\affiliation{Canadian Institute for Advanced Research,Toronto, M5G 1Z8 Canada.}

\date{{\today}}
\begin{abstract}
We observe that the diffusive spin current in a strongly interacting degenerate Fermi gas of $^{40}$K precesses about the local magnetization. As predicted by Leggett and Rice,
precession is observed both in the Ramsey phase of a spin-echo sequence, and in the nonlinearity of the magnetization decay. At unitarity, we measure a Leggett-Rice parameter $\gamma = 1.08(9)$ and a bare transverse spin diffusivity $D_0^\perp = 2.3(4)\,\hbar/m$ for a normal-state gas initialized with full polarization and at one fifth of the Fermi temperature, where $m$ is the atomic mass. One might expect $\gamma = 0$ at unitarity, where two-body scattering is purely dissipative. We observe $\gamma \rightarrow 0$ as temperature is increased towards the Fermi temperature, consistent with calculations that show the degenerate Fermi sea restores a non-zero $\gamma$. Tuning the scattering length $a$, we find that a sign change in $\gamma$ occurs in the range $0 < (k_F a)^{-1} \lesssim 1.3$, where $k_F$ is the Fermi momentum. We discuss how $\gamma$ reveals the effective interaction strength of the gas, such that the sign change in $\gamma$ indicates a switching of branch, between a repulsive and an attractive Fermi gas.
\end{abstract}

\maketitle


Transport properties of unitary Fermi gases have been studied extensively in the past few years. Due to strong inter-particle interactions at unitarity, various transport coefficients like viscosity and spin diffusivity are bounded \cite{Thomas2010,Zwierlein2011,*Sommer:2011ig,Bardon:2014} by a conjectured quantum minimum \cite{Kovtun:2005tz,*Schafer:2009vf,*Enss:2011,Bruun:2011ue,Enss:2012b}, in three dimensions. On the other hand, transport in two-dimensional unitary Fermi gases shows anomalous behavior, apparently violating a quantum limit \cite{Kohl2013}. This remains to be understood.

\begin{figure}[bt!]
\includegraphics[width= .9\columnwidth]{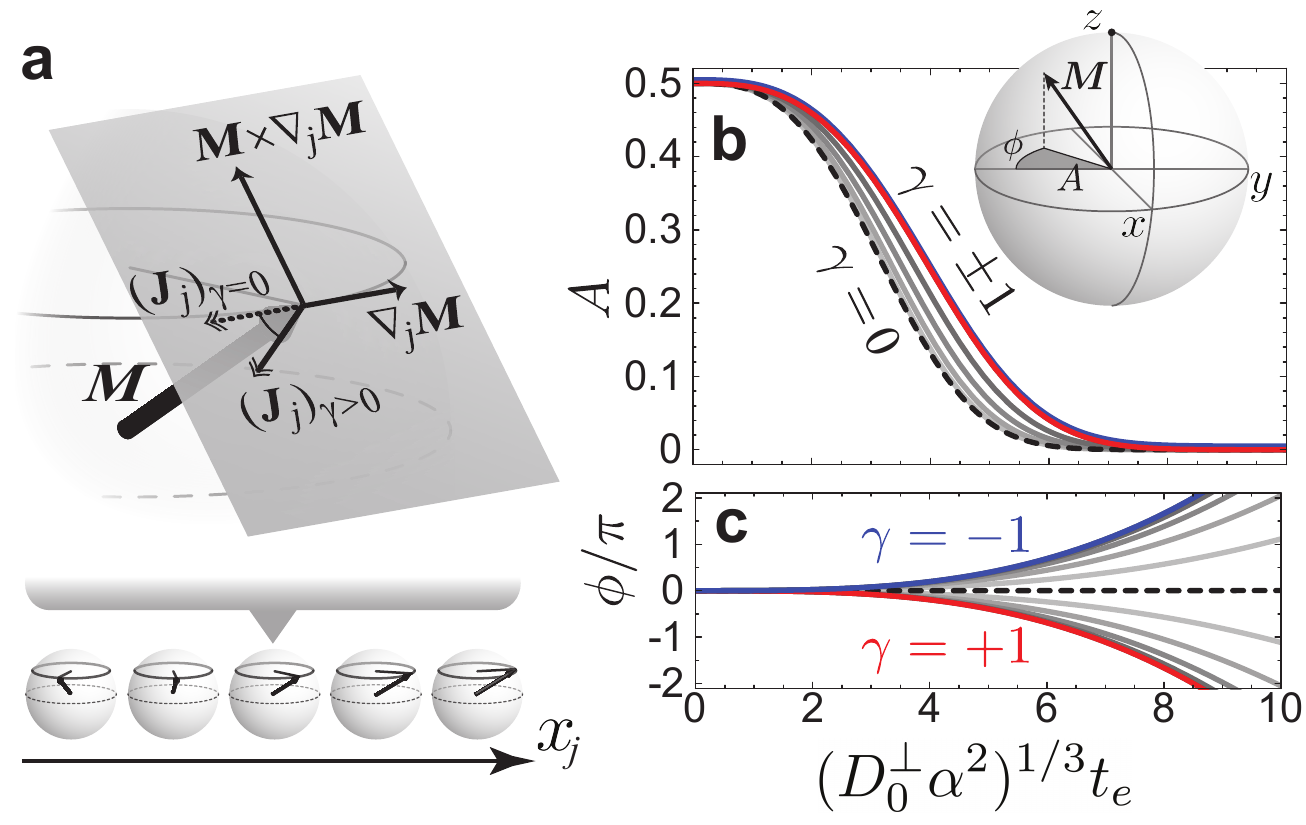}
\caption{The Leggett-Rice effect. {(a)} In a transverse spin spiral along $x_j$, the gradient $\nabla_j \svec{M} \perp \svec{M}$ drives a spin current $\svec{J}_j \perp \svec{M}$, as described by Eq.~(1). For $\gamma \neq 0$, $\svec{J}_j$ is rotated around $M$ by $\arctan(\gamma)$ compared to $(J_j)_{\gamma = 0}$. In a spin-echo experiment, this causes both a slower decay of amplitude, $A = |M_{x}+iM_y|$ shown in {(b)}, as well as an accumulated phase, $\phi = -\arg(i M_{x}-M_y)$ shown in {(c)}. The case of $\theta=5\pi/6$ and full initial polarization is plotted. Dashed lines in (b) and (c) show $\gamma=0$, and gray lines show steps of $0.2$ up to $\gamma=\pm 1$.
\label{fig:schematic}}
\end{figure}

In the case of spin diffusion, experiments so far \cite{Zwierlein2011,*Sommer:2011ig,Kohl2013,Bardon:2014} have been interpreted with a spin current proportional to the magnetization gradient, $\svec{J}_j = - D \grad \svec{M}$, where $D$ is the diffusion constant \cite{Hahn:1950tv,*Carr:1954wf,*Torrey:1956tk}, and $\svec{M} = \langle M_x, M_y, M_z \rangle$ is the local magnetization. Bold letters indicate vectors in Bloch space and the subscript $j \in \{1,2,3\}$ denotes spatial direction. In general, $\svec{J}_j$ has both a longitudinal component $\svec{J}_j^{\parallel}\parallel \svec{M}$ and a transverse component $\svec{J}_j^{\perp}\perp \svec{M}$. Longitudinal spin currents are purely dissipative, and the standard diffusion equation applies~\cite{Jeon:1989,*Mullin:1992wi,*Meyerovich:1985pla,*Meyerovich:1994tk,Bruun:2011ue,Levin:2011,*Bruun:2011hn,*Heiselberg:2012vd,*Duine:2012,*Huse:2012,*Chevy:2012,*Enss:2012a,*Goulko:2013jb,Enss:2012b}. However, as Leggett and Rice pointed out~\cite{LR:1968,*Leggett:1970}, the transverse spin current follows
\begin{equation} \label{eq:J}
\svec{J}_j^{\perp}  = - D_{\rm eff}^\perp \grad \svec{M} - \gamma \svec{M} \times D_{\rm eff}^\perp \grad \svec{M},
\end{equation}
where $D_{\rm eff}^\perp = D_0^\perp/(1+\gamma^2 M^2)$ is the effective transverse diffusivity and $\gamma$ is the Leggett-Rice (LR) parameter~\footnote{The relation of $\gamma$ to the conventional LR parameter $\mu$ is $\gamma = n \mu/2$.} (see Fig.\,\ref{fig:schematic}a). Physically, the second term describes a reactive component of the spin current that precesses around the local magnetization. This precession has been observed in weakly interacting Fermi gases~\cite{Du:2008de,*Du:2009hm,Heinze:2013ko,Kohl2013} and is a manifestation of the so-called identical spin-rotation effect~\cite{Piechon:2009cr,*Natu:2009gt}, which is intimately related to the LR effect~\cite{Miyake:1985hz}. In a unitary Fermi gas, however, neither the existence of the LR effect nor the value of $\gamma$ has been measured. In this Letter, we provide the first evidence for LR effects in a unitary Fermi gas, and measure $\gamma$ using a spin-echo technique.

Our experiments are carried out in a trapped cloud of $^{40}$K atoms using the two lowest-energy Zeeman states $\updown$ of the electronic ground-state manifold \cite{SM}. Interactions between these states are tuned by the Feshbach resonance \cite{RevModPhys2010} at $202.1$\,G. We start with a completely spin-polarized sample in the lowest-energy state $\down$. This large initial polarization enhances the LR effect, since $\gamma$ appears as a product with $M$ in the equations of motion. Polarization also suppresses the critical temperature of superfluidity \cite{Chandrasekhar:1962,*Clogston:1962,Shin:2008db} to below the temperature range explored here.

We probe magnetization dynamics using a series of radio-frequency (rf) pulses. At time $t=0$, a resonant pulse with area $\theta$ creates a superposition of $\down$ and $\up$, in which $M_z = -\cos(\theta)$ and $M_{xy} \equiv M_x + i M_y = i\sin(\theta)$. During time evolution, a controlled magnetic-field gradient $B' = 17.0(7)$\,G/cm oriented along the $x_3$ spatial direction leads to a variation of the phase of the superposition, twisting the $xy$-magnetization $M_{xy}$ into a spiral. After a spin-refocusing pulse ($\pi_x$) at time $t_\pi$, the spiral untwists, and all spins realign at the echo time $t_e = 2t_\pi$. A readout $\pi/2$ pulse with variable phase lag closes an internal-state interferometer, which is observed using time-of-flight imaging after Stern-Gerlach state separation \cite{SM}. The contrast and phase of interference fringes measure the trap-averaged values of both the amplitude $A = |M_{xy}|$ and phase $\phi = -\arg (i M_{xy})$ of the $xy$-magnetization at the echo time.

In such an echo experiment, $M_z$ is manipulated only by rf pulses, and is otherwise conserved globally and even locally in a uniform system. However the gradient of $M_{xy}$ initializes irreversible spin currents that cause the transverse magnetization to decay. The resulting dynamics are described by~\cite{LR:1968,*Leggett:1970}
\begin{equation} \label{eq:Mperp}
	\partial_t M_{xy} = -i \alpha x_3 M_{xy} + D_{\rm eff}^\perp (1+i \gamma M_z) \nabla^2_3 M_{xy} 
\end{equation}
where $\alpha = B' \Delta \mu / \hbar$, and $\Delta \mu$ is the difference in magnetic moment between $\up$ and $\down$.

If $\gamma=0$, the solution of Eq.\,(\ref{eq:Mperp}) is $A(t_e) = A_0 \exp{(- D_0^\perp \alpha^2 t_e^3/12)}$  \cite{Hahn:1950tv,*Carr:1954wf,*Torrey:1956tk}. For $\gamma \neq 0$, but for small $A_0$ or short times, 
$A(t_e)$ decays with the same functional form but with $D_0^\perp$ replaced by $D_{\rm eff}^{\perp} < D_0^\perp$. At longer times, however, the magnetization loss can differ significantly from this simple form. The full solution to Eq.\,(\ref{eq:Mperp}) is given by
\begin{eqnarray}
\label{eq:Asol}
	A(t_e) &=& A_0 \sqrt{\frac{1}{\eta}\, \mathcal{W}\left(\eta \exp{\left[\eta-\frac{D_0^\perp \alpha^2 t_e^3}{6(1 + \gamma^2 M_z^2)}\right]}\right)}\,,\\
\label{eq:Qsol}
	\phi(t_e) &=& \gamma M_z \ln \left(\frac{A(t_e)}{A_0}\right)\,,
\end{eqnarray}
where $\eta = \gamma^2A_0^2 / (1 + \gamma^2 M_z^2)$ and $\mathcal{W}(z)$ is the Lambert-W function. Figure~\ref{fig:schematic}(b,c) shows typical plots of Eqs.~(\ref{eq:Asol}) and (\ref{eq:Qsol}) for a variety of $\gamma$. The LR effect is seen in both amplitude and phase dynamics. However, $A(t_e)$ alone is an ambiguous signature. For example, a similar shape of $A(t_e)$ could result from a magnetization dependence of $D_0^\perp$, as is predicted to occur below the so-called anisotropy temperature due to restrictions of collisional phase space \cite{Jeon:1989,*Mullin:1992wi,*Meyerovich:1985pla,*Meyerovich:1994tk,Enss:2013ti}. The evolution of $\phi(t_e)$ and its relation to $A(t_e)$, on the other hand, are unique features of the LR effect that are sensitive to the signs of $\gamma$ and $M_z$.

\begin{figure}[tb]
\includegraphics[width=\columnwidth]{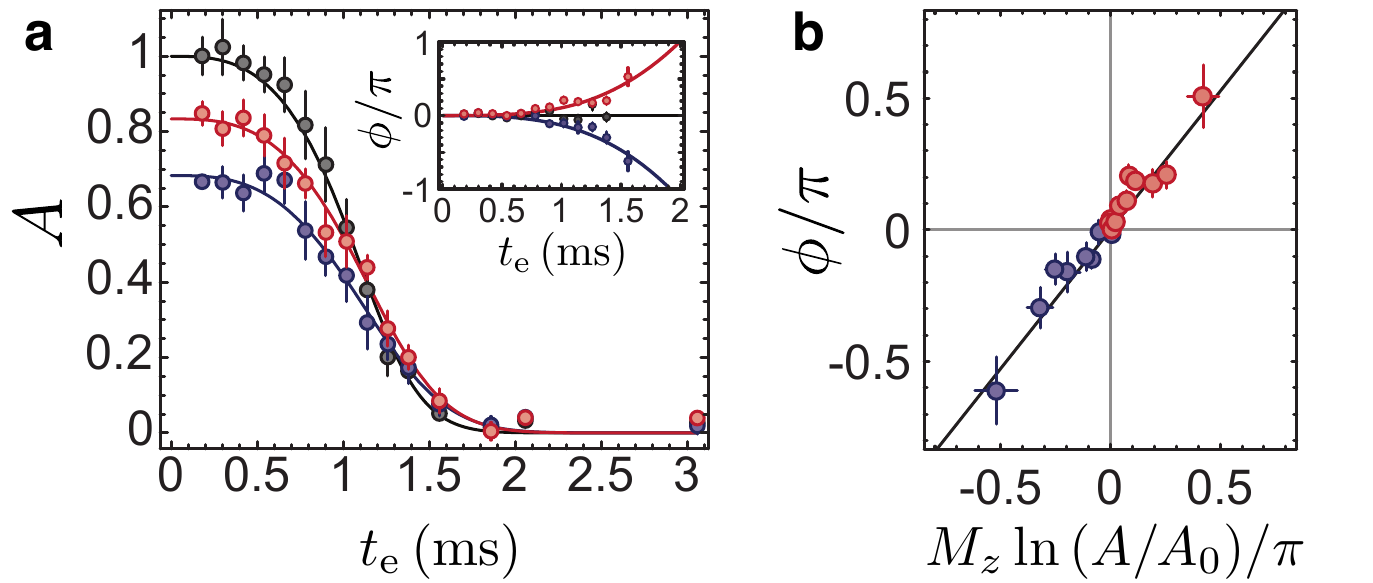}
\caption{ {(a)} Amplitude $A$ and phase $\phi$ (inset) of the $xy$ magnetization measured at unitarity for $\TTFi\simeq 0.2$ and with $M_z = 0.00(5)$ (black circles), $M_z =0.74(2)$ (blue) and $M_z =-0.54(3)$ (red). All data is taken at a spin-echo time. Error bars represent uncertainties from the fit to a full interferometric fringe.
{(b)} Plot of $\phi(t_e)$ vs. $M_z \ln{[A(t_e)/A_0]}$ for the two cases where $M_z \neq 0$. The solid line is a linear fit to both data sets that is used to extract $\gamma$. Error bars represent combined uncertainties from fit and extrapolation.
The solid lines in (a) represent fits with Eq.\,(\ref{eq:Asol}) using the value of $\gamma$ obtained by the analysis presented in (b).
\label{fig:LRE}}
\end{figure}

Figure~\ref{fig:LRE}a shows the measured $A(t_e)$ and $\phi(t_e)$ at unitarity and initial temperature $\TTFi \simeq 0.2$, for three initial-pulse areas, where $T_F$ is the Fermi temperature. We test for the LR effect by plotting $\phi(t_e)$ as a function of $M_z \ln[A(t_e)/A_0]$  (see Fig.\,\ref{fig:LRE}b), where $A_0$ is obtained by extrapolating $A(t_e)$ to $t_e=0$ and full initial polarization is assumed, i.e.\ $M_z^2 + A_0^2 = 1$. We determine $\gamma$ from a linear fit to the data plotted as in Fig.\,\ref{fig:LRE}b, following Eq.\,(\ref{eq:Qsol}). Fixing $\gamma$, diffusivity is determined from a subsequent fit of Eq.\,(\ref{eq:Asol}) to $A(t_e)$ (lines in Fig.\,\ref{fig:LRE}a). From this analysis, we obtain $\gamma = 1.08(9)$ and $D_0^\perp = 2.3(4)\,\hbar/m$ at unitarity. These best-fit transport coefficients should be regarded as an average over the trapped ensemble, and over the full range of magnetization. Furthermore, during the spin diffusion process, temperature rises due to demagnetization \cite{Bardon:2014,Ragan:2002tx}. Full demagnetization at unitarity increases $T/T_F$ from $0.2$ to about $0.4$, but the temperature rises less for smaller pulse angles and for weaker interactions.

At low temperature, Landau Fermi liquid (LFL) theory \cite{BaymPethick} provides a microscopic interpretation of these transport parameters: $D_0^\perp=2 \chi_0 \tau_\perp \epsilon_F/(3m^* \chi)$, where $\tau_\perp$ is the transport lifetime, 
$\epsilon_F=(\hbar k_F)^2/2m$ is the local Fermi energy of an ideal gas, $\chi$ is the magnetic susceptibility and $\chi_0$ is its ideal-gas value, and $m^*$ is the effective mass.  The LR parameter is $\gamma = -(4 \chi_0/3 \chi) (\tau_\perp  \epsilon_F/\hbar) \lambda$, with $\lambda\equiv -\hbar\gamma/(2m^* D^{\perp}_0)$ a dimensionless coefficient.  The thermodynamic response of the system is parameterized by LFL parameters $F_{0,1}^{a,s}$: $m^*=m(1+F^s_1/3)$, $\chi m/m^*= \chi_0/(1+F^a_0)$, and $\lambda = (1+F^a_0)^{-1} - (1 + F^a_1/3)^{-1}$~\cite{LR:1968,*Leggett:1970,BaymPethick}. However, $\tau_\perp$ is accessible only by transport measurements. If we use the $D_0^\perp$ at our lowest probed temperature, with $\chi/\chi_0=0.73$ and $m^*/m=1.13$ from a thermodynamic measurement~\cite{Nascimbene:2010en,*Nascimbene:2011fn}, we estimate $\tau_\perp \approx 2.5 \hbar/\epsilon_F$. Since LFL theory assumes $\tau_\perp \gg \hbar/\epsilon_F$, the transport lifetime we find is at or near the lower self-consistent bound for a quasiparticle treatment \cite{Deng:2013cs}.

Notice that $\lambda$ has two contributions: $F^a_0$, corresponding to the effective magnetic field produced by local magnetization, and $F^a_1$, corresponding to a spin vector potential created by a local spin current. The latter has no analogue for weakly interacting fermions. A spin-echo experiment such as ours can constrain $F^a_1$, if all other LFL parameters are known. We find $\lambda \approx -0.2$ at unitarity, smaller in magnitude than $2.1 \leq \lambda \leq 2.7$ in liquid $^3$He \cite{Greywall:1983dg}.  Combined with $F^a_0=1.1(1)$ from thermodynamic measurements~\cite{Nascimbene:2011fn}, this implies $F^a_1 \approx 0.5$ for a unitary Fermi gas. Repeating our measurements at smaller magnetization and lower temperature would provide a test of LFL theory for a unitary gas. For instance, our estimated value of $F^a_1$ is near the upper limit to be consistent with $F^s_1=0.4(1)$ determined from $m^*$ in a balanced gas \cite{Nascimbene:2010en,*Nascimbene:2011fn}, since LFL theory requires $F^a_1 < F^s_1$ \cite{Leggett:1968ta}. 

\begin{figure}[t!]
\includegraphics[width=0.9\columnwidth]{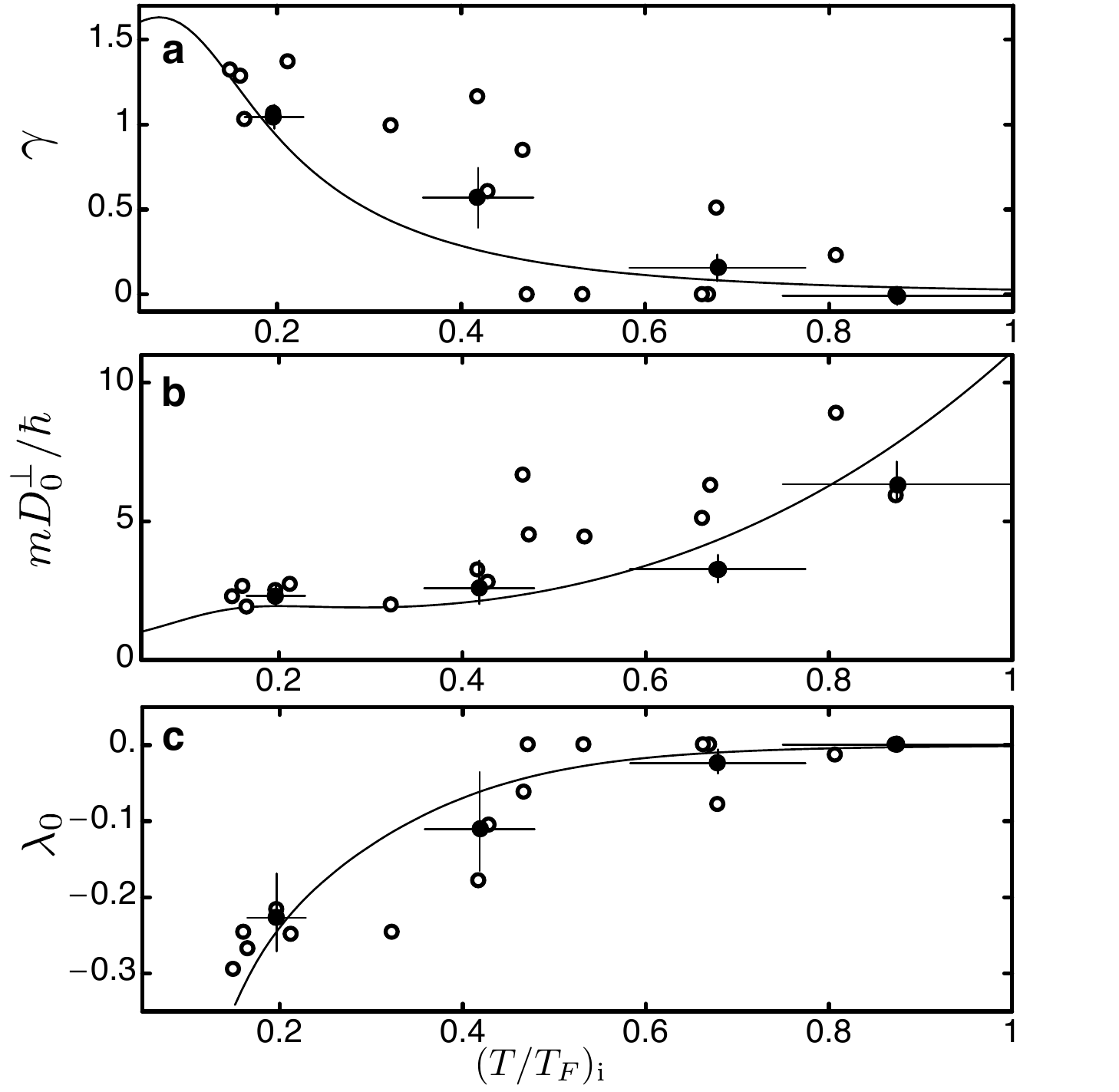}
\caption{Spin transport at unitarity. {(a)} The measured LR parameter $\gamma$, {(b)} diffusivity $D_0^\perp$, and {(c)} the ratio $\lambda_0= -\hbar \gamma/(2 m D_0^\perp)$ are shown versus the initial reduced temperature $\TTFi$. Solid points are each from a phase-sensitive measurement as shown in Fig.\,\ref{fig:LRE}. Horizontal and vertical error bars represent statistical and fit uncertainties. For these data, $N$ ranges from $50(5) \times 10^3$ at low temperature to $18(4) \times 10^3$ at high temperature. Open circles are results from a fit of Eq.\,(\ref{eq:Asol}) to $\theta \simeq \pi/2$ data such as the black circles in Fig.\,\ref{fig:LRE}a, and also to data from Ref.\,\onlinecite{Bardon:2014}. Here, we fix $M_z$ and vary $\gamma$ (chosen {\it a posteriori} to be nonnegative) and $D_0^\perp$.
Although the two methods provide similar values on average, the phase-sensitive measurements provide reduced scatter for $\gamma \lesssim 0.5$, and are sensitive to the sign of $\gamma$. Solid lines show a kinetic theory calculation in the limit of large imbalance, and using the local reduced temperature at peak density \cite{SM}.
\label{fig:unitarity}}
\end{figure}

\begin{figure}[t!]
\includegraphics[width=0.9\columnwidth]{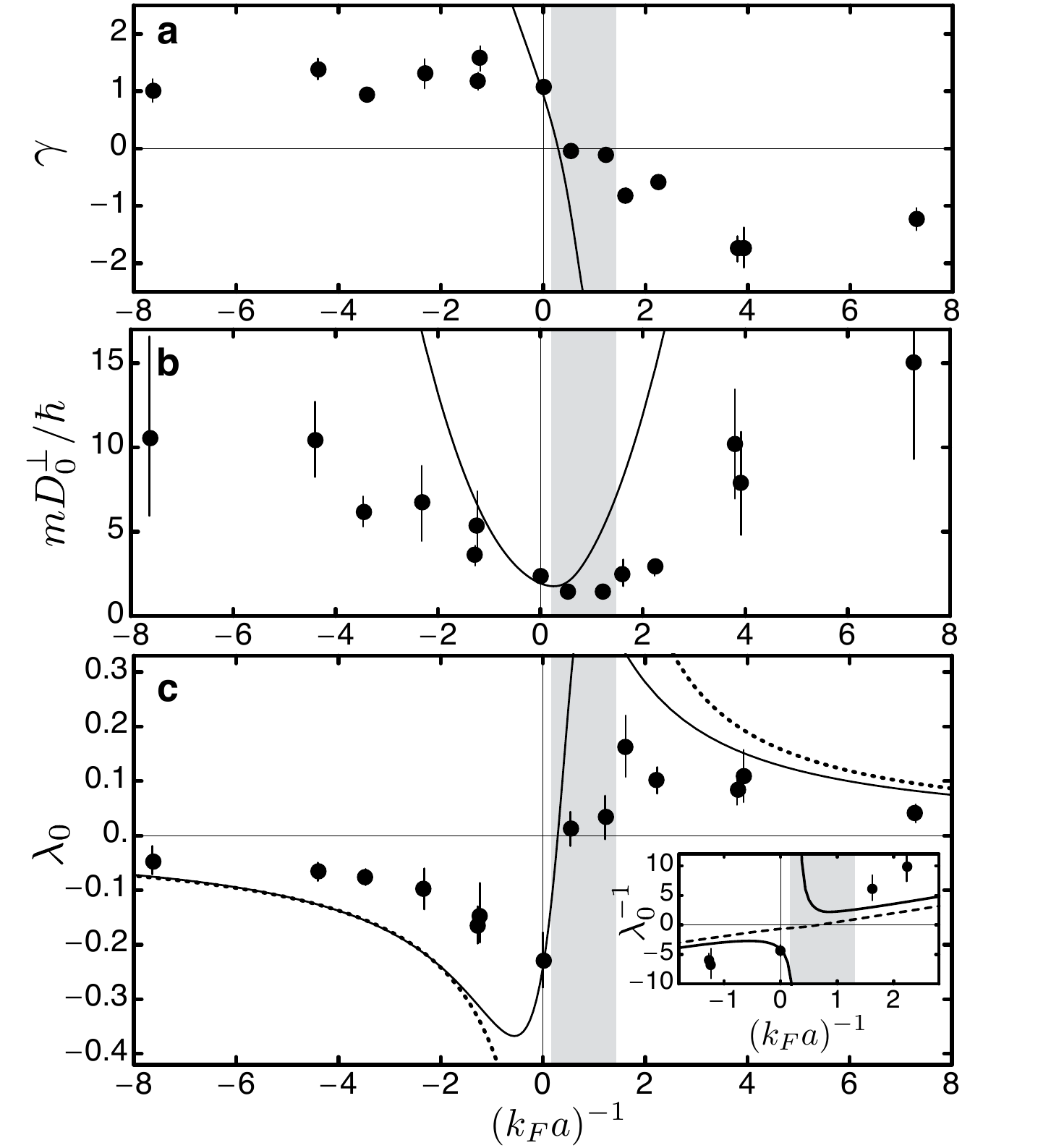}
\caption{Effect of interaction strength on spin transport.
{(a)} LR parameter $\gamma$, {(b)} $D_0^\perp$, and {(c)} $\lambda_0$ as a function of $(k_F a)^{-1}$. The error bars represent fit uncertainties. For these data, $\TTFi = 0.18(4)$ and $N=40(10) \times 10^3$, where uncertainty is due to number variation between runs. In the range $0 < (k_F a)^{-1} \lesssim 1.3$ (indicated in grey) both free atoms and Feshbach dimers are present, as discussed in the text and in Fig.\,\ref{fig:mols}. Solid lines show a kinetic theory calculation \cite{SM} at $\TTFi=0.20$; the dotted line in (c) shows the weakly interacting limit $\lambda_0 = [\pi/(2k_Fa)-1]^{-1}$ for a balanced $T=0$ gas \cite{Miyake:1985hz}. The inset to (c) shows $\lambda_0^{-1}$, and includes a calculation using the momentum averaged upper branch T-matrix (solid line) as well as $\lambda^{-1}_0 =  (4\epsilon_F/3n){\cal{T}}^{-1}(\rvec{0},0)$. The sign change of $\lambda_0$ at $0.4 \leq (k_F a)^{-1} \leq 1$ is a robust feature of theory, and is consistent with our data.
\label{fig:vskFa} }
\end{figure}

Figures \ref{fig:unitarity} and \ref{fig:vskFa} show how spin transport depends on temperature and interaction strength. We reinterpret our earlier work \cite{Bardon:2014} to have observed the effective diffusivity $D_{\rm eff}^\perp$; whereas here we find both $\gamma$ and the bare $D_0^\perp$. Within the range of parameters explored, $D_0^\perp$ is still consistent with the conjectured limit \cite{Kovtun:2005tz,*Schafer:2009vf,*Enss:2011,Bruun:2011ue,Enss:2012b}.

We compare our data to a kinetic theory \cite{Enss:2013ti,SM} in which collisions are described in terms of the many-body T-matrix $\T(\bq,\omega)$, which gives the low-energy effective interaction between fermions near the Fermi surface, whose center-of-mass momentum and energy are $\hbar \bq$ and $\hbar \omega$, respectively \cite{SM}. Kinetic theory relates $\gamma$ to a momentum average of $\T$~\cite{Mullin:1992wi,Enss:2013ti,SM}; this result is shown in  Figs.~\ref{fig:unitarity} and \ref{fig:vskFa}.  At low temperatures, $\T$ is peaked about $\rvec{q}=0,\omega=0$, and $\gamma$ is well approximated by $\gamma=-\Re \T(\rvec{0},0) \tau_\perp n/ \hbar$, where $n$ is number density \cite{Jeon:1989}. We use this to interpret some of our results in what follows. A simple interpretation of the $\T$ is given by its weakly interacting limit in vacuum, $\T \rightarrow -(4 \pi \hbar^2/m) f(k)$, where $f(k)=-1/(a^{-1}+ik)$ is the s-wave scattering amplitude, $a$ is the s-wave scattering length, and $k$ is the relative wave vector of two colliding fermions. More generally, the sign of $\Re\T$ reveals whether dressed interactions in the gas are attractive or repulsive \cite{SM}.

The conceptual simplicity of $\lambda$ is that the ratio $-\gamma/D_0^\perp$ eliminates $\tau_\perp$, leaving a quantity proportional to $\Re\T$. However, $m^*$ is not known for the full range of polarizations, temperatures, and interaction strengths probed here. We report instead $\lambda_0 \equiv -\hbar \gamma/(2 m D_0^\perp) \propto \Re \T(\rvec{0},0)$ with the bare mass
\footnote{For theory curves, we use $\chi m^*/m \chi_0=1$, which is correct in the weakly interacting limit but introduces a systematic error for an interacting gas, on the order of 20\% for the balanced, low-temperature unitary gas.}. The pair $D_0^\perp$ and $\lambda_0$ encapsulate the dissipative and reactive effects of scattering.

At unitarity, we observe that $\lambda_0$ depends sensitively on $\TTFi$ and approaches zero at high temperatures (Fig.\,\ref{fig:unitarity}c). This is in contrast to the temperature insensitivity of spin-wave behavior in a weakly interacting Fermi gas \cite{Du:2008de,*Du:2009hm}. At high temperatures, $\T$ reduces to the two-body scattering amplitude mentioned above, which is purely imaginary at unitarity. As a result, $\lambda_0$ approaches zero. At low temperature, however, the degenerate Fermi sea restores a non-zero $\Re \T$ and hence $\lambda_0$.

For all interaction strengths in Fig.\,\ref{fig:vskFa}, data are analyzed as described above for unitarity. However, the validity of our hydrodynamic model likely breaks down at weaker interactions. We estimate that the mean free path $\ell \approx 3 D_0 /k_F$ at peak density changes from 300\,nm at $(k_F a)^{-1} \approx 0$ to 3\,$\mu$m at $|k_F a|^{-1} = 2$. This approaches both the pitch of the spin-spiral, $1/\alpha t \approx 4$\,$\mu$m at $t_e \sim 1$\,ms, and the Thomas-Fermi radius of the cloud, 5\,$\mu$m, along the $x_3$ direction. Thus we expect the data analysis based on Eq.\,(\ref{eq:J}) to be most accurate in the strongly interacting regime.

Figure~\ref{fig:vskFa}a shows an approximately linear change in $\gamma$ across $-1 \leq (k_F a)^{-1} \leq 3$. This agrees only qualitatively with our kinetic calculation (solid line). However the calculation is for full and constant polarization, and does not encompass the dynamic temperature, nor the inhomogeneous density of the cloud. A second salient feature of the data is the minimum in $D_0^\perp$ near the scattering resonance, which is reminiscent of behavior seen in other transport parameters \cite{Zwierlein2011,Kohl2013,Elliott:2014bv,*Bluhm:2014ts}. Strong collisions impede the transport of spin. As with $\gamma$, the best-fit $D_0^\perp$ saturates at larger $|k_F a|^{-1}$, perhaps due to finite-size effects that remain to be understood.

The LR effect changes sign in the range $0 < (k_F a)^{-1} \lesssim 1.3$ (see Figs.\,\ref{fig:vskFa}a,c). This indicates that the effective interaction between fermions changes sign as one tunes the system across the Feshbach resonance~\cite{SM}. Such a sign change is only possible if the system switches from the ``upper branch'' of the energy spectrum near the Feshbach resonance \cite{Pricoupenko:2004gm,Pilati:2010jd,Chang04012011,Shenoy:2011dm} to the lower branch, in which interactions are attractive.

\begin{figure}[tb]
\includegraphics[width= \columnwidth]{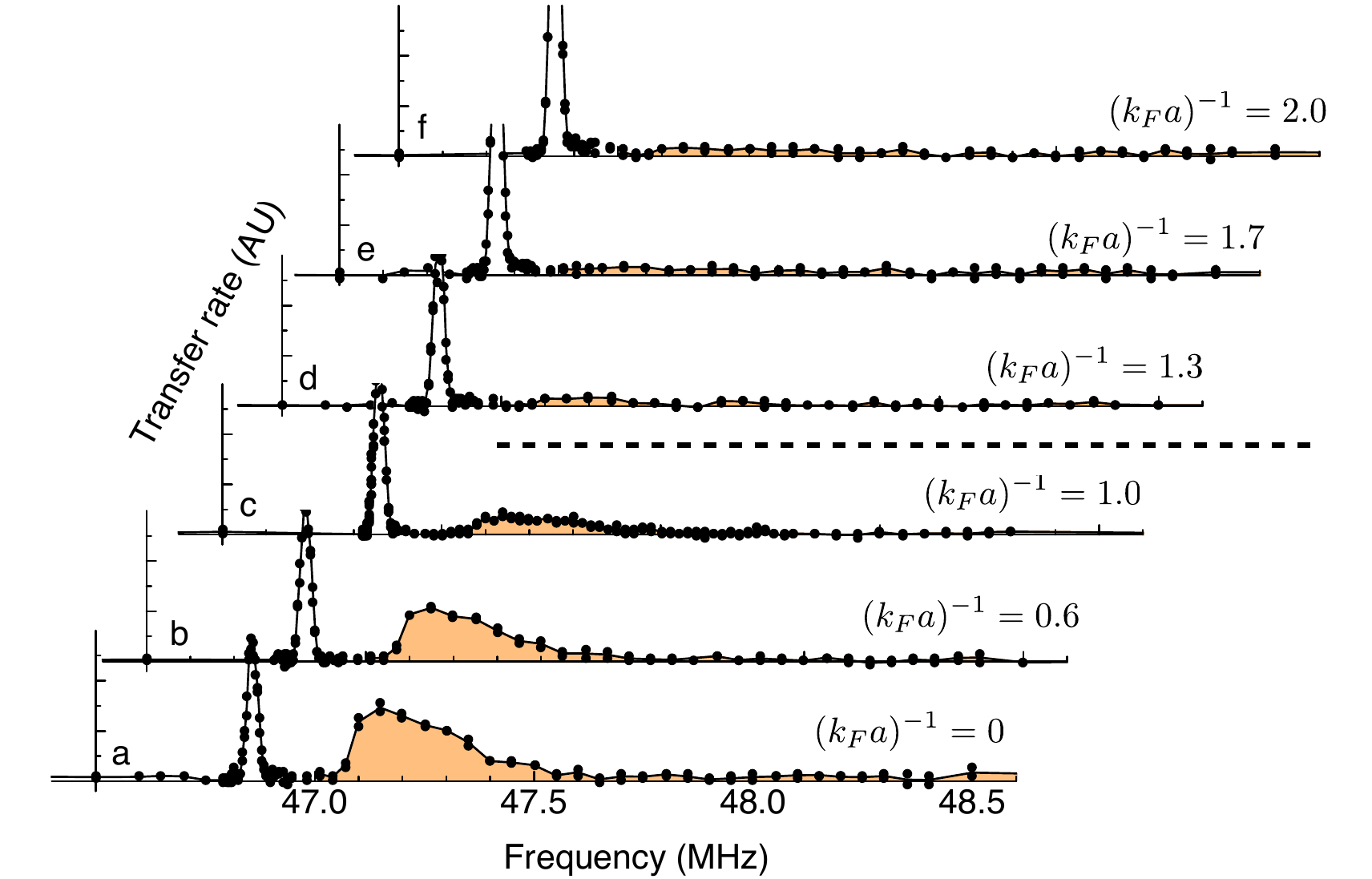}
\caption{Presence of dimers above the Feshbach resonance. At the indicated initial $(k_F a)^{-1}$, a superposition is created with $\theta=\pi/2$ and held for 3\,ms. The field is then swept to 200.0\,G in 5\,ms, which magneto-associates some lower-branch pairs into dimers with a binding energy of $h\times200$\,kHz. Dimers are identified with their rf dissociation spectrum, using an 80-$\mu$s pulse near the 46.85\,MHz spin-flip resonance from the $\up$ state to a previously unoccupied Zeeman state \cite{[{See }] [{ and further discussion in Supplementary Material.}]Regal:2003ex}. Each plot shows the transfer rate versus rf frequency. For traces (a), (b), and (c), there is a clearly identified molecular feature above $47.0$\,MHz (spectral weight shaded in orange). However for traces (d), (e), and (f), the spectral weight above $47.0$\,MHz is insensitive to field, and consistent with the noise of the measurement.
\label{fig:mols} }
\end{figure}

The sign change of ${\rm Re}\T(\rvec{0},0)$ has been previously discussed \cite{Pekker:2011fh,Sodemann:2012eq} in the context of an upper-branch instability, in which atoms decay to form bound pairs in the lower branch \cite{Sanner:2012gf,*Lee:2012jm,Zhang:2011ga,Cui:2010ej,*Conduit:2010cw,*Ma:2012kk,*Pilati:2014cb}. To search for dimers that would be produced by the pairing instability, we use a combination of magneto-association and spin-flip spectroscopy (Fig.\,\ref{fig:mols}). We observe that for $0 < (k_F a)^{-1} \lesssim 1.3$, the same range of $(k_F a)^{-1}$ where $\gamma$ changes sign, there are weakly-bound Feshbach dimers, even though $a>0$ for the entire experimental sequence. We shade this range in Fig.\,\ref{fig:vskFa}, to flag the simultaneous presence of upper- and lower-branch atoms. No clear evidence of Feshbach dimers appears at $(k_F a)^{-1} \geq 1.3$, however more deeply bound dimers would not appear in our detection method \cite{Zhang:2011ga}. 

In summary, we have observed an unambiguous signature of the Leggett-Rice effect in a strongly interacting Fermi gas. In the limit of zero temperature, $\gamma$ and $D_0^\perp$ are scale-invariant universal transport parameters of the unitary Fermi gas. The value of $D_0^\perp$ reveals the strength of dissipative scattering in the gas. It is near the proposed quantum limit, such that the inferred value of $\tau_\perp$ is comparable to the ``Planck time'' $\hbar/\epsilon_F$ \cite{Hartnoll:2014vb,*Bruin:2013hc}. This raises the possibility that incoherent transport may play a role, i.e., that a quasiparticle-based picture may be incomplete.

The Leggett-Rice effect reveals the reactive component of scattering between fermions of unlike spin. The nonzero value of $\gamma$ tells us that spin waves in a unitary Fermi gas are dispersive \cite{Silin:1958tc}, or in other words, that the gas has a spin stiffness in the long-wavelength limit \cite{Mineev:2005js,DemlerPC}. Spin stiffness is an essential ingredient of ground-state magnetic textures \cite{PhysRevB.79.224403,*PhysRevA.80.013607}. Even though magnetic ordering does not occur in the conditions of our experiments, this same energetic term is clearly observed with our interferometric measurement. \nocite{Akimoto:2003vs,Enss12,deMelo93s}

\acknowledgments{We thank E.\ Demler, R.\ Ragan, and A.\ Paramekanti  for stimulating conversations, and N.\ Zuber for experimental assistance. This work was supported by NSERC, by AFOSR under FA9550-13-1-0063, by ARO under W911NF-14-1-0282, and by RGC under HKU-709313P.}


%

\section{Supplementary Materials}

\paragraph*{Spin current in the hydrodynamic limit.}

The local magnetization density $\M$ obeys a continuity equation 
\be \label{eqSM:Mcont}
\partial_t \bm{\mathcal{M}} + \svec\Omega_0 \times \M   = -\grad \Jj \ee
where $\svec\Omega_0 = \langle 0, 0, \Omega_0 \rangle$ with $\Omega_0$ the Larmour frequency due to an external field, and $\Jj$ is the spin current density.
In Eq.\,(\ref{eqSM:Mcont}) and below, the $j$ subscript indicates a spatial direction, $\partial_t$ is a time derivative, and bold quantities denote vectors in Bloch space. The magnetization density is a vector whose magnitude is $|\M|= (n_\uparrow - n_\downarrow)/2$. Whereas $\M$ is conserved in the frame rotating with the external field, the spin current is in general not conserved. In the hydrodynamic limit, the local steady-state spin current obeys
\be \Jj + \Jj \times \mu \M + D_0 \grad \M = 0 \ee
whose solution is  
\be \label{eqSM:J}
\Jj  = - D_\mathrm{eff} [ \grad \M -  \mu \M \! \times \! \grad \M - \mu^2 (\M \cdot \grad \M)\M ] \ee
where $D_\mathrm{eff} = D_0/(1+\mu^2 \mathcal{M}^2)$. If $\mu=0$, (\ref{eqSM:J}) reduces to ``static diffusion'', described by Fick's law, $\Jj  = - D_0 \grad \M$.

\paragraph*{Transverse and longitudinal spin currents.}

The general expression for the current, Eq.\,(\ref{eqSM:J}), can be broken into two components. Writing $\M = \mathcal{M} \svec{m}$ with $\mathcal{M} = |\M|$, 
\be \grad \M = (\grad \mathcal{M}) \svec{m} + ( \grad \svec{m} ) \mathcal{M}. \ee
The first term is parallel to $\M$, and the second term is perpendicular in Bloch space to $\M$, since it is the derivative of a unit vector.

For gradients $\grad \M$ that are {\em parallel} to $\M$, the resultant current is called {\em longitudinal}. Here, the second term in Eq.\,(\ref{eqSM:J}) is zero, the third term becomes proportional to $\mu^2 \mathcal{M}^2 \grad \M$, and thus
\be \label{eqSM:Jlong}
\Jj^{||} = - D^{||}_0 \grad \M \ee
where the $\mu$ dependence has cancelled out, and only the bare longitudinal diffusivity is left.

For gradients $\grad \M$ that are {\em perpendicular} to $\M$, the resultant current is called {\em transverse}. In this case, the third term in Eq.\,(\ref{eqSM:J}) is zero, and we have
\be \label{eq:Jtrans}
\Jj^\perp = - D^\perp_\mathrm{eff} \left[ \grad \M + \mu \M \times \grad \M \right] \ee
where $D^\perp_\mathrm{eff} = D^\perp_0/(1+\mu^2 \mathcal{M}^2)$. Since at low temperature diffusivity can differ between the longitudinal and transverse cases, they are labelled $D_0^{||}$ and $D_0^\perp$ respectively. In Landau Fermi liquid theory \cite{BaymPethick}, $D_0^{||}$ and other damping coefficients such as viscosity and conductivity scale as $\sim T^{-2}$ at low temperature, while $D_0^\perp$ remains anomalously finite \cite{Jeon:1989,*Mullin:1992wi,Meyerovich:1985pla,*Meyerovich:1994tk,Akimoto:2003vs}. Also note that  Eq.\,(\ref{eqSM:Jlong}) is independent of $\mu$, so the transverse spin current alone is sensitive to the LR effect.

\paragraph*{Transverse polarization current.}

Magnetization is observed through the density-weighted trap average of dimensionless polarization $\svec M=2\M/n$ with $|\svec{M}| \leq 1$. Paired with $\svec M$ is a polarization current $\svec{J}_j = 2\Jj/n$. Treating only the transverse component, (\ref{eq:Jtrans}) becomes
\be \label{eqSM:Jperp}
\svec{J}_j^\perp = - D^\perp_\mathrm{eff} \, \grad  \svec M - D^\perp_\mathrm{eff} \gamma \svec M \times  \grad \svec M \ee
where we have ignored spatial gradients of density, and $\gamma \equiv \mu n / 2$, such that
$\gamma \svec{M} = \mu \M$. Equation~(\ref{eqSM:Jperp}) is Eq.\,(1) in the main text.

Now, consider the continuity equation (\ref{eqSM:Mcont}). Assuming a static density profile, $\partial_t n =0$ and ignoring spatial gradients of density,
\be \label{eq:Pcont}
 \partial_t \svec{M} + \svec\Omega_0 \times \svec{M} = -\grad  \svec{J}_j^\perp \ee
where repeated indices are summed.

\paragraph*{Experimental methods.}

Fermionic, spin-polarized $^{40}$K atoms are cooled sympathetically with bosonic $^{87}$Rb atoms. Initially both species are trapped in a microfabricated magnetic trap, where $^{87}$Rb is evaporated directly. A subsequent stage of evaporative cooling is performed in a crossed-beam optical dipole trap, with $^{40}$K atoms in the $|{f}=9/2, m_f=-9/2\rangle$ state and $^{87}$Rb atoms in the $|{f}=1, m_f=1\rangle$ state. Here, $f$ and $m_f$ denote the total angular momentum and the corresponding magnetic quantum number, respectively. At the end of cooling, residual $^{87}$Rb atoms are removed with a resonant light pulse typically leaving $N=4\times10^4$ $^{40}$K atoms. The trap has a mean trapping frequency $\bar{\omega}/2\pi=470 (20)$ Hz and an aspect ratio of 4:1:1. The Feshbach field $B_3$ and the magnetic field gradient $\nabla_{3}B_3$ are applied along a tight axis of the trap. The $|{-z}\rangle$, $|{+z}\rangle$, and rf-probe states in the main text refer to the high-field states adiabatically connected to the low-field $m_f$ = -9/2, -7/2, and -5/2 states of the $f=9/2$ hyperfine manifold of the electronic ground state.

The initial temperature is determined by imaging the density distribution after time-of-flight expansion, and fitting it to a Fermi-Dirac distribution. For this measurement, the gas is fully polarized in the $|{f}=9/2,m_f=-9/2\rangle$ state so no interaction corrections are required. For the data presented in Fig.\,4, $N= 40(10) \times10^4$ atoms with an initial temperature of $250 (40)$\,nK. The initial Fermi energy is then $E_{F,\mathrm{i}}/h=29(4)$\,kHz, such that $\TTFi=0.18(4)$, where $T_F \equiv E_F/k_B$.
The global (trap-wide) Fermi energy $E_F$ is subtly different from the local Fermi energy $\epsilon_F = \hbar^2 k_F^2 /2m$ with a local $k_{F,\mathrm{pol}} \equiv (6 \pi^2 n)^{1/3}$ in a polarized gas and $k_{F,\mathrm{mix}} =  k_{F,\mathrm{pol}}/2^{1/3}$ in an unpolarized gas. In both cases, $n$ is the local total number density.
In Fig.~4 and in theory discussion, $k_F \equiv k_{F,\mathrm{mix}}$, as is conventional. However, we use the set of definitions for the polarized gas to define the initial reduced temperature $(T/T_F)_{\rm i}$. In order to tune the initial temperature for data presented in Fig.\,3, we vary the loading and evaporation sequence, affecting both the absolute temperature of the gas and the total atom number. The uncertainties stated for atom numbers and temperatures are a combination of statistical and calibration uncertainties.

Since the theory to which we compare our experimental results are based on uniform models, we match the local reduced temperature $k_B T/\epsilon_{F,\mathrm{pol}}$, where $\epsilon_{F,\mathrm{pol}} = \hbar^2 k_{F,\mathrm{pol}}^2 /2m$. At the center of the trap, this local reduced temperature is minimal, since the density is highest. The peak density is given by $n_p = \lambda_T^{-3} f_{3/2}(z_p)$, where $\lambda_T = \sqrt{2 \pi \hbar^2/m k_B T}$ is the thermal de Broglie wavelength, $f_{3/2}$ is the  statistical function for a uniform Fermi gas, and $z_p$ is the local fugacity at peak density. The fugacity (and chemical potential) are constrained by the temperature and total particle number per spin component, according to $6 f_3(z_p) = (k_B T/E_{F,\mathrm{i}})^{-3}$.

During depolarization, energy and entropy increase, changing the temperature. As discussed in \cite{Bardon:2014}, this `intrinsic heating' effect can be calculated at unitarity using an experimentally measured density equation of state. At low temperature, the effect is strongest: a $(T/T_F)_\mathrm{i}=0$ cloud will heat to $(T/T_F)_\mathrm{f} \approx 0.35$, and at our lowest $(T/T_F)_\mathrm{i} \approx 0.20$, the final reduced temperature is $(T/T_F)_\mathrm{f} \approx 0.40$.  At high temperature, the released entropy and interaction energy becomes negligible, and the effect vanishes. The rise in $T$ is also reduced for smaller $A_0$.

We measure the polarization decay for various initial magnitudes of the transverse polarization $|M_{xy}|=\sin{\left(\theta\right)}$ using a $\theta-\pi-\pi/2$ pulse sequence. The first and second pulse have the same phase, but the final pulse has a variable relative phase lag. Varying this phase lag reveals the magnitude $A=|M_{xy}|$ and phase $\phi=-\arg \left(i M_{xy}\right)$ of the transverse polarization in the oscillation of the relative population with. The amplitude and phase of $M_{xy}$ is then determined from a sinusoidal fit to this oscillation. We are sensitive to the relative frequency between the drive $\omega$ and the atomic frequency $\omega_0$, with a precision of roughly $1/t$, where $t$ is the hold time. For $t\geq 1.5$ ms, we find that our field stability (roughly 1 kHz, or a few parts in $10^5$) is insufficient to preserve a reproducible relative phase, resulting in a randomized phase for long hold times. To avoid this, we choose a magnetic field gradient such that the timescale for depolarization is smaller than this coherence time.

The analysis technique developed utilizes both the amplitude and phase data for three separate mixing angles per value of $\gamma$ and $D_0^{\perp}$ reported. The mixing angles used are $\theta\approx0.32\pi$, $\theta\approx0.50\pi$, and $\theta\approx0.74\pi$ such that the pulse area in the rf sequence is varied simply by changing the pulse duration after optimizing a $\pi$ pulse. To extract $\gamma$ and $D_0^{\perp}$, we first fit the amplitude data $A(t)$ for each mixing angle using an exponential decay with a free exponent, $A_0\exp{\left[-(t/\tau)^\eta\right]}$, to extrapolate the initial amplitude $A_0$. Typically, this amplitude is slightly different from the desired value due to imperfect pulse area. For the mixing angles where $A_0\neq1$, assuming $M_z^2+A_0^2=1$, we rescale the phase data and plot $\phi(t)$ as a function of $M_z\log{\left[A(t)/A_0\right]}$. A linear fit to this rescaled data provides a single value of $\gamma$ for both mixing angles. Fixing $\gamma$, we then fit the amplitude of the data set where $M_z\approx 0$ using Eq. (3) to obtain an initial guess for $D_0^{\perp}$. Using this value as a starting point we fit the amplitude data for all three mixing angles using Eq. (3) and minimize the residuals to extract a single $D_0^{\perp}$. This method is used to extract the values of $\gamma$ and $D_0^{\perp}$ in Fig.\,4 and the closed circles in Fig.\,3.

The open circles in Fig.\,3 are extracted from the amplitude data alone. In this case, only the $M_z \approx 0$ mixing angle is used (we have included reanalyzed data from \cite{Bardon:2014}). A single fit of the amplitude data for this mixing angle to Eq. (3) provides both $|\gamma|$ and $D_0^{\perp}$. However, this fit is extremely sensitive to slight nonlinearities in the amplitude and often fails for data in which $\gamma\leq0.5$. The phase-sensitive method greatly reduces scatter, and is sensitive to the sign of $\gamma$.

We control the magnetic field and its gradients through a combination of magnetic field coils and micro-fabricated wires on an atom chip located about 200 $\mu$m from the atoms. We tune the field $|\mathbf{B}|=B_{3}$ near 202.10(2) G, at which the two states $|\pm{z}\rangle$ undergo a Feshbach resonance. The field is calibrated by measuring the $|-z\rangle$ to $|+z\rangle$ transition frequency and converting the frequency to magnetic field through the Breit-Rabi formula. During a measurement at a single value of $\left(k_{F}a\right)^{-1}$ the field drifts by as much as 0.02 G which translates to a systematic uncertainty of $\pm0.02$ in $\left(k_{F}a\right)^{-1}$ at unitarity. We control the field gradients $\nabla_2B_3$ and $\nabla_3B_3$ by adjusting the sum and difference of small currents through parallel chip wires near the atoms, setting $\nabla_2B_3=0$. We calibrate the gradients by repeating spectroscopy measurements on a cloud translated by piezo-actuated mirrors on the trapping beams.

Our imaging scheme allows us to simultaneously count the populations of atoms in states $|\pm{z}\rangle$. This is achieved with a Stern-Gerlach pulse to separate the trapped spin states, rf state manipulation during time of flight in a gradient, and finally, resonant absorption imaging of the $|f=9/2, m_f=-9/2\rangle$ to $|f=11/2, m_f=-11/2\rangle$ cycling transition. Imaging occurs after jumping the magnetic field to 209 G, the zero crossing of the s-wave scattering resonance, to minimize interaction effects during time of flight.

\paragraph*{Kinetic theory.}

The primary theoretical calculation to which we compare our experimental data (solid lines in Fig.\,3 and Fig.\,4) uses the Boltzmann equation to find the non-equilibrium time evolution of the spin distribution function in response to the applied magnetic field gradient. Collisions between fermions of unlike spin use the many-body T-matrix computed in the medium of surrounding fermions [see \onlinecite{Enss:2013ti} and below].  This corresponds to the Nozi\`eres-Schmidt-Rink approximation used to compute the Fermi-liquid parameters, and it is justified as the leading-order term in a systematic large-N expansion in the number of fermion flavors \cite{Enss12}. The calculation does not include any finite-size effects. We calculate transport parameters in the limit of large imbalance, which corresponds to initial conditions, and suppresses superfluidity in the calculation.

We have solved the kinetic theory for a homogeneous, fully polarized
Fermi gas.  Previous work has shown that for temperatures $0.3 T_F \lesssim
T \lesssim T_F$, the transverse diffusivity $D^\perp_0 \sim 2\hbar/m$ is robust and does not differ much from the longitudinal diffusivity
$D_\parallel$ \cite{Enss:2013ti}. From the transverse scattering time $\tau_\perp$ (see Eq.\,(53) in Ref.\,\onlinecite{Enss:2013ti}) we compute both 
\be D^{\perp}_0 = \frac{\tau_{\perp}}{2{\cal{M}}}\int \frac{d^3\rvec{k}}{(2\pi)^3}\sum_i v_{ki}v_{kj}(f_{\rvec{k}\uparrow}-f_{\rvec{k}\downarrow})\ee
 and the spin-rotation parameter $\gamma$.  The latter is a weighted momentum average of the many-body T matrix $\T(\bq,\omega)$~\cite{Enss:2013ti}:
\begin{align}\gamma =& -\frac{n\tau^2_{\perp}}{4\hbar D^{\perp}_{0}{\cal{M}}^2}\int \frac{d^3\rvec{k}_1}{(2\pi)^3}\frac{d^3\rvec{k}_2}{(2\pi)^3}v_{1j}(v_{1j}-v_{2j})(f_{1\uparrow}-f_{1\downarrow})\nonumber\\&\times(f_{2\uparrow}-f_{2\downarrow})\mathrm{Re}{\cal{T}}(\rvec{k}_1+\rvec{k}_2,\xi_{1\uparrow}+\xi_{2\downarrow}).\label{gamma_m}\end{align}
Here $1,2$ are shorthand for $\rvec{k}_1, \rvec{k}_2$, $f_{\rvec{k}\sigma} \equiv [\exp(\beta\xi_{\rvec{k}\sigma})+1]^{-1}$ is the Fermi distribution for $\xi_{\rvec{k}\sigma}\equiv (\hbar\rvec{k})^2/2m - \mu_{\sigma}$, and $v_{\rvec{k}j}$ is the $j$th Cartesian component of the velocity.     
This result is used to calculate $\gamma$ and the ``momentum-averaged'' $\lambda_0\equiv -\hbar\gamma/(2mD^{\perp}_0)$ shown in Figs.\,(3) and (4) in the main text.  

We relate our numerical results for the homogeneous system to the
measurements in the trapping potential using the local-density
approximation.  The experimental response is dominated by the
center of the trap with the highest local density, and in Fig.\,3 we
show the diffusivity of a homogeneous system of that same density and
reduced temperature, as described above. The spin-rotation
parameter $\gamma$ thus estimated agrees well with the experiment at
unitarity.  Away from unitarity, at weak coupling $|k_Fa|\to0$, $\T(\rvec{0},0)\propto a$ (see below), giving $\gamma\sim -1/a$ for a
homogeneous system, and $\lambda_0 \sim a$. For $\lambda_0$, the geometry dependent diffusive scattering time $\tau_\perp$ drops out, and we find qualitative agreement between the homogeneous and trapped systems.

\paragraph*{The $\T$-matrix on the upper and lower branches.} 

In the simplest ladder approximation~\cite{deMelo93s},
\begin{center}
\includegraphics[width=0.94\linewidth]{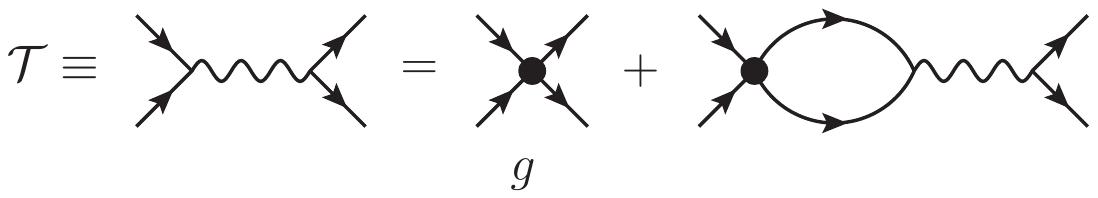}
\end{center}
where $g=4\pi \hbar^2 a/m$ is the bare interaction vertex and the straight lines are non-interacting Fermi Green's functions.  This leads to
 \begin{align}\T^{-1}(\bq,\omega) &=\frac{m}{4\pi \hbar^2}\left[\frac{1}{a} + \frac{i}{\hbar}\sqrt{m\left(\hbar \omega+\mu_{\uparrow}+\mu_{\downarrow}-\tfrac{\hbar^2 q^2}{4m}\right)}\right]\nonumber\\&+\int \frac{d^3\rvec{k}}{(2\pi)^3}\frac{f_{\rvec{k}\uparrow}+f_{\rvec{k}-\rvec{q} \downarrow}}{\hbar \omega-\xi_{\rvec{k}\uparrow}-\xi_{\rvec{k}-\rvec{q}\downarrow}}.\label{ppvertex}\end{align}

Physically, $\T(\rvec{q},\omega)$ can be thought of as the effective interaction between a pair of spin $\uparrow$ and $\downarrow$ fermions close to the Fermi surface, with centre-of-mass momentum $\rvec{q}$ and energy $\hbar \omega$ with respect to the sum $\mu_{\uparrow}+\mu_{\downarrow}$ of the chemical potentials.  It is renormalized from the bare $s$-wave interaction $g$ by many-body effects.  

$\T(\rvec{q},\omega)$ is strongly peaked about $\rvec{q}=0,\omega=0$. Replacing it by this value, (\ref{gamma_m}) reduces to~\cite{Jeon:1989}
\be \gamma =  -\frac{\T(\rvec 0, 0) n\tau_{\perp}}{\hbar}.\label{gamma}\ee
We have confirmed that (\ref{gamma}) provides a good approximation to (\ref{gamma_m}), in particular the location of the zero-crossing which indicates the onset of a pairing instability (see below).  (\ref{gamma_m}) remains finite at this point, however, whereas (\ref{gamma}) diverges.  Equation (\ref{gamma}) thus means that that the LR parameter is sensitive to the effective interaction between spin $\uparrow$ and $\downarrow$ fermions.  Moreover, the fact that $\tau_{\perp}>0$ means that the sign of $\gamma$ determines whether the effective interaction is attractive [$\T(\vzero,0)<0$] or repulsive [$\T(\vzero,0)>0$].  A change in the sign of $\gamma$ [$\T(\vzero,0)$] from negative to positive [positive to negative] as some parameter is tuned indicates the onset of a pairing instability.  

The equilibrium state of the system is characterized by chemical potentials $\mu_{\sigma}$ obeying the thermodynamic condition~\cite{deMelo93s} \be n_{\sigma} = \sum_{\bk}f_{\bk,\sigma} + \frac{\partial}{\partial\mu_{\sigma}} \frac{1}{\beta}\sum_{\bq,\nu_m}\ln \T (\bq,i\nu_m). \label{NSR}\ee
Using these values in (\ref{ppvertex}), the effective interaction $\T(\vzero,0)$ is attractive throughout the entire BCS--BEC crossover and, at low enough temperatures,  fermions are paired up.  We emphasize that this is true despite the fact that the bare interaction vertex changes sign at unitarity.   

At the same time, the effective interaction need not be everywhere attractive in the excited ``upper branch'' state~\cite{Chang04012011,Shenoy:2011dm}.  In this state, the fermions are unbound, in scattering states.  In the limit $k_Fa\to 0^+$, the chemical potentials are given by their ideal gas values (when $T\ll T_F$) $\mu_{\sigma}= \hbar^2(6\pi^2 n_{\sigma})^{2/3}/2m$, and are not negative, as happens in this limit on the lower branch (for a spin balanced mixture), where the ground state is a Bose-Einstein condensate of dimer molecules.  With these ideal gas values of the chemical potentials, (\ref{ppvertex}) reduces to the expected ``hard-sphere'' result~\cite{Chang04012011} $\T(\vzero,0) \to 4\pi \hbar^2 a/m>0$ in the $k_Fa\to 0^+$ limit of the upper branch, corresponding to a negative value for the LR parameter.  

To characterize this excited metastable state outside the weak-coupling limit $k_Fa\to 0^+$, we solve (\ref{NSR}) self-consistently, but remove the isolated molecular pole in the T-matrix to determine the upper branch chemical potentials~\cite{Shenoy:2011dm}. In this way, we exclude the possibility of the formation of bound states, which would correspond to the equilibrium state with dimer molecules when $k_Fa>0$. Using the resulting values for $\mu_{\sigma}$ at $T=0.5T_{F}$ and $M_z = (n_{\uparrow}-n_{\downarrow})/n=0.25$, we find that $\T(\vzero,0)$ diverges in the vicinity of $(k_Fa)^{-1} \simeq 0.8$, becoming negative for smaller values.  As noted above, this sign change indicates the onset of an instability of the upper branch towards the formation of pairs and the resulting evolution to the equilibrium lower branch~\cite{Sodemann:2012eq,Pekker:2011fh}.  In agreement with previous work~\cite{Sodemann:2012eq}, the critical value at which this happens depends very weakly on $M_z$.  

Ignoring LFL corrections, $D^{\perp}_0 = (2\epsilon_F/3m)\tau_{\perp}$ and $\lambda_0 = -\hbar\gamma/(2mD^{\perp}_0)$.  Combining these with (\ref{gamma}) gives
\be \lambda^{-1}_0 \equiv \frac{4\epsilon_F}{3n}\T^{-1}(\vzero,0).\ee
In the inset of Fig.\,4c in the main text, we show the upper branch value of $\lambda_0^{-1}$. Its zero-crossing is consistent with both the experimental values and the sign change of $\lambda_0$ found with the momentum-averaged T-matrix (\ref{gamma_m}).

\end{document}